\documentclass[twocolumn,showpacs,aps,prd]{revtex4-1}

\usepackage{amssymb}
\usepackage{graphicx}
\usepackage{dcolumn}
\usepackage{bm}
\usepackage{epsfig}
\usepackage{amsfonts}
\usepackage{keyval,graphicx}
\usepackage{textcomp,wasysym}
\usepackage{subfigure}
\usepackage{hyphenat}
\usepackage{xcolor}

\long\def\inst#1{\par\nobreak\kern 4pt\nobreak
    {\itshape #1}\par\vskip 10pt plus 3pt minus 3pt}
\RequirePackage{xspace}
\usepackage{relsize}

\begin{document}

\newcommand{\chisq}[1]{$\chi^{2}_{#1}$}
\newcommand{\etap}{\eta^{\prime}}
\newcommand{\etac}{\eta_c}
\newcommand{\etacp}{\eta_c(2S)}
\newcommand{\pip}{\pi^{+}}
\newcommand{\pim}{\pi^{-}}
\newcommand{\piz}{\pi^{0}}
\newcommand{\rhoz}{\rho^{0}}
\newcommand{\az}{a_{0}(980)}
\newcommand{\fz}{f_{0}(980)}
\newcommand{\pipm}{\pi^{\pm}}
\newcommand{\psip}{\psi^{\prime}}
\newcommand{\psipp}{\psi(3770)}
\newcommand{\jpsi}{J/\psi}
\newcommand{\ar}{\rightarrow}
\newcommand{\GeV}{GeV/$c^2$}
\newcommand{\MeV}{MeV/$c^2$}
\newcommand{\br}[1]{\mathcal{B}(#1)}
\newcommand{\cinst}[2]{$^{\mathrm{#1}}$~#2\par}
\newcommand{\crefi}[1]{$^{\mathrm{#1}}$}
\newcommand{\crefii}[2]{$^{\mathrm{#1,#2}}$}
\newcommand{\crefiii}[3]{$^{\mathrm{#1,#2,#3}}$}
\newcommand{\HRule}{\rule{0.5\linewidth}{0.5mm}}

\title{\Large \boldmath \bf Search for the radiative transitions 
$\psipp\to\gamma\etac$ and $\gamma\etacp$}

\author{
\small
M.~Ablikim$^{1}$, M.~N.~Achasov$^{8,a}$, X.~C.~Ai$^{1}$, O.~Albayrak$^{4}$, M.~Albrecht$^{3}$, D.~J.~Ambrose$^{41}$, F.~F.~An$^{1}$, Q.~An$^{42}$, J.~Z.~Bai$^{1}$, R.~Baldini Ferroli$^{19A}$
% ...
\\
\vspace{0.2cm}
(BESIII Collaboration)\\
\vspace{0.2cm} {\it
$^{1}$ Institute of High Energy Physics, Beijing 100049, People's Republic of China\\
% ...
$^{3}$ Bochum Ruhr-University, D-44780 Bochum, Germany\\
$^{4}$ Carnegie Mellon University, Pittsburgh, Pennsylvania 15213, USA\\
% ...
$^{8}$ G.I. Budker Institute of Nuclear Physics SB RAS (BINP), Novosibirsk 630090, Russia\\
% ...
$^{19}$ (A)INFN Laboratori Nazionali di Frascati, I-00044, Frascati, Italy; (B)INFN and University of Perugia, I-06100, Perugia, Italy\\
% ...
$^{41}$ University of Rochester, Rochester, New York 14627, USA\\
$^{42}$ University of Science and Technology of China, Hefei 230026, People's Republic of China\\
% ...
$^{a}$ Also at the Novosibirsk State University, Novosibirsk, 630090, Russia\\
% ...
}}

\vspace{0.4cm}

\date{\today}

\begin{abstract} 
By using a 2.92 fb$^{-1}$ data sample taken at $\sqrt{s} = 3.773$ GeV with the BESIII detector operating at the BEPCII collider,  we search for the radiative transitions $\psipp\to\gamma\etac$ and $\gamma\etacp$ through the hadronic decays  $\etac(\etacp)\to K^0_SK^\pm\pi^\mp$. No significant excess of signal events above background is observed. We set upper limits at a 90\% confidence level for the product branching fractions to be $\mathcal{B}(\psipp\to\gamma\etac)\times \mathcal{B}(\etac\to K^0_SK^\pm\pi^\mp) < 1.6\times10^{-5}$ and $\mathcal{B}(\psipp\to\gamma\etacp)\times \mathcal{B}(\etacp\to K^0_SK^\pm\pi^\mp) < 5.6\times10^{-6}$. Combining our result with world-average values of $\mathcal{B}(\etac(\etacp)\to K^0_SK^\pm\pi^\mp)$, we find the branching fractions $\mathcal{B}(\psipp\to\gamma\etac) < 6.8\times10^{-4}$ and $\mathcal{B}(\psipp\to\gamma\etacp) < 2.0\times10^{-3}$ at a 90\% confidence level. 
\end{abstract}


\maketitle


\section{Introduction} 
\label{sec:intro} 

The nature of the excited $J^{PC}=1^{--}$ $c\bar{c}$ bound states above the $D\bar{D}$ threshold is of interest but still not well known. The $\psipp$ resonance, as the lightest charmonium state lying above the open charm threshold, is generally assigned to be a dominant $1^3D_1$ momentum eigenstate with a small $2^3S_1$ admixture~\cite{Rosner}. It has been thought almost entirely to decay to $D\bar{D}$ final states~\cite{prl39,prl40}. Unexpectedly, the BES Collaboration found a large inclusive non-$D\bar{D}$ branching fraction, $(14.7\pm3.2)\%$, by utilizing various methods~\cite{Abli1,Abli2,Abli3,Abli4}, neglecting interference effects, and assuming that only one $\psipp$ resonance exists in the center-of-mass energy between 3.70 and 3.87~GeV. A later work by the CLEO Collaboration, taking into account the interference between the resonance decays and continuum annihilation of $e^+e^-$, found a contradictory non-$D\bar{D}$ branching fraction, $(-3.3\pm1.4^{+6.6}_{-4.8})\%$~\cite{CLEO-nonDD}. The BES results suggest substantial non-$D\bar{D}$ decays, although the CLEO result finds otherwise. In the exclusive analyses, the BES Collaboration observed the first hadronic non-$D\bar{D}$ decay mode, $\psipp\to\jpsi\pip\pim$~\cite{pipijpsi}. Thereafter, the CLEO Collaboration confirmed the BES observation~\cite{nondd-cleo}, and observed other hadronic transitions, including  $\piz\piz\jpsi$, $\eta\jpsi$~\cite{nondd-cleo}, the E1 radiative transitions $\gamma\chi_{cJ}(J=0,1)$~\cite{gchi1,gchi2}, and the decay to light hadrons $\phi\eta$~\cite{phieta}. While experimentalists have been continuing to search for exclusive non-$D\bar{D}$ decays of the $\psipp$, the sum of the observed non-$D\bar{D}$ exclusive components still makes up less than 2\% of all decays~\cite{pdg}, which motivates the search for other exclusive non-$D\bar{D}$ final states.

The radiative transitions $\psipp\to\gamma\etac(\etacp)$ are supposed to be highly suppressed by selection rules, considering the $\psipp$ is predominantly the $1^3D_1$ state. However, due to the non-vanishing photon energy in the decay, higher multipoles beyond the leading one could contribute~\cite{IML}. Recently, authors of Ref.~\cite{IML} calculated the partial decay widths  $\Gamma(\psi(3770)\to\gamma\eta_c)=(17.14^{+22.93}_{-12.03})$~keV and $\Gamma(\psi(3770)\to\gamma\eta_c(2S))=(1.82^{+1.95}_{-1.19})$~keV (with corresponding branching fractions $\mathcal{B}(\psipp\to\gamma\etac)=(6.3^{+8.4}_{-4.4})\times10^{-4}$ and  $\mathcal{B}(\psipp\to\gamma\etacp)=(6.7^{+7.2}_{-4.4})\times10^{-5}$ calculated with $\Gamma_{\psipp}=27.2\pm1.0$~MeV~\cite{pdg}) by taking into consideration significant contributions from the intermediate meson loop (IML) mechanism, which is important for exclusive transitions, especially when the mass of the initial state is close to the open channel threshold. Experimental measurements of the branching fractions $\mathcal{B}(\psipp\to\gamma\etac(\etacp))$ will be very helpful  for testing theoretical predictions and providing further constraints on the IML contributions.

In this paper, we present the results of searches for the radiative transitions $\psipp\to\gamma\etac(\etacp)$. In order to avoid high combinatorial background and to get good resolution, the $\etac(\etacp)$ is reconstructed in the most widely used hadronic decay $\etac(\etacp)\to K^0_SK^\pm\pi^\mp$, which contains only charged particles and has a large branching fraction. As a cross-check, the branching fraction of the E1 transition $\psipp\to\gamma\chi_{c1}$ is also measured using the decay mode $\chi_{c1}\to K^0_SK^\pm\pi^\mp$. The results reported in this paper are based on a 2.92~fb$^{-1}$ data sample taken at $\sqrt{s} = 3.773$~GeV, accumulated by the BESIII detector operating at the BEPCII $e^+e^-$ collider.

\section{The BESIII Experiment and Monte Carlo simulation}
\label{sec:detector}

The BESIII detector~\cite{besnim} (operating at the BEPCII accelerator) is a major upgrade of the BESII detector (which operated at the BEPC accelerator) and it is used for the study of physics in the $\tau$-charm energy region~\cite{taucharm}. The design peak luminosity of the double-ring $e^+e^-$ collider, \nohyphens{BEPCII}, is $10^{33}$ cm$^{-2}$s$^{-1}$ at a beam current of 0.93~A. The BESIII detector has a geometrical acceptance of 93\% of $4\pi$ and consists of four main components: (1) A small-celled, main drift chamber (MDC) with 43~layers, which provides measurements of ionization energy loss~($dE/dx$) and charged particle tracking. The average single wire resolution is 135~$\mu$m, and the momentum resolution for charged particles with momenta of 1~GeV/$c$ in a 1~T magnetic field is 0.5\%. (2) An electromagnetic calorimeter~(EMC), which is made of 6240~CsI~(Tl) crystals arranged in a cylindrical shape~(barrel) plus two end caps. For 1.0~GeV photons, the energy resolution is 2.5\% in the barrel and 5\% in the end caps, and the position resolution is 6~mm in the barrel and 9~mm in the end caps. (3) A time-of-flight system~(TOF), which is used for particle identification~(PID). It is composed of a barrel part made of two layers with 88~pieces of 5~cm thick and 2.4~m long plastic scintillators in each layer, and two end caps with 96~fan-shaped, 5~cm thick plastic scintillators in each end cap. The time resolution is 80~ps in the barrel, and 110~ps in the end caps, corresponding to a 2$\sigma$ K/$\pi$ separation for momenta up to about 1.0~GeV/$c$. (4) A muon chamber system, which consists of 1272~m$^2$ of resistive plate chambers arranged in 9~layers in the barrel and 8~layers in the end caps and is incorporated in the return iron of the super-conducting magnet. The position resolution is about 2~cm.

Monte Carlo~(MC) simulations of the full detector are used to determine the detection efficiency of each channel, to optimize event-selection criteria and to estimate physics backgrounds. The {\sc geant4}-based~\cite{GEANT4} simulation software, BESIII Object Oriented Simulation ~\cite{BOOST}, contains the detector  geometry and material description, the detector response and signal digitization models, as well as records of the detector  running conditions and performance. The production of the $\psipp$ resonance is simulated with the MC event generator {\sc kkmc}~\cite{kkmc1,kkmc2}, which includes initial-state radiation~(ISR). The signal channels are generated with the expected angular distributions for $\psipp\to\gamma\etac,~\gamma\etacp,~\gamma\chi_{c1}$. The subsequent $\etac,~\etacp,~\chi_{c1}\to K^0_SK^\pm\pi^\mp$ are produced according to measured Dalitz plot distributions, which are obtained from the processes $\psi(3686)\to\gamma\etac(\chi_{c1})\to\gamma K^0_SK^\pm\pi^\mp$ for $\etac(\chi_{c1})\to K^0_SK^\pm\pi^\mp$ and $B^{\pm}\to K^\pm\etacp\to K^\pm(K^0_SK\pi)^0$ for $\etacp\to K^0_SK^\pm\pi^\mp$, as  measured by the Belle Collaboration~\cite{Belle-etacp}. To investigate possible background contaminations, MC samples of $\psipp$ inclusive decays equivalent to 10~times that of the data, and $e^+e^-\to\gamma_{\rm ISR}\jpsi$, $\gamma_{\rm ISR}\psi(3686)$, $q\bar{q}$ ($q=u,~d,~s$) equivalent to 5~times that of  the data are generated. The decays are generated with {\sc evtgen}~\cite{evtgen} for the known decay modes with branching fractions taken from the Particle Data  Group (PDG)~\cite{pdg} or by the Lundcharm model {\sc lundcharm}~\cite{lund} for the unmeasured decays.

\section{Event Selection} 
\label{sec:selection} 

Each charged track except those from $K^0_S$ decays is required 
to be within 1~cm in the radial 
direction and 10~cm along the beam direction consistent with the run-by-run-determined  
interaction point. The tracks must be within the MDC fiducial volume, 
$|\cos\theta| < 0.93$, where $\theta$ is the polar angle with respect to  
the $e^+$ beam direction. Charged-particle identification~(PID) is 
based on combining the $dE/dx$ and TOF information  to form the 
variable $\chi^2_{\rm PID}(i)=(\frac{dE/dx_{\rm measured}-dE/dx_{\rm expected}}
{\sigma_{dE/dx}})^2 +(\frac{{\rm TOF}_{\rm measured}-{\rm TOF}_{\rm expected}}
{\sigma_{\rm TOF}})^2$.  The values $\chi^2_{\rm PID}(i)$ are calculated 
for each charged track for each particle hypothesis $i$ 
($i=$ pion, kaon, or proton). 

Photon candidates are reconstructed by clustering EMC crystal energies. 
The energy deposited in the nearby TOF scintillator is included to improve 
the reconstruction efficiency and the energy resolution. Showers in the 
EMC must satisfy fiducial and shower-quality requirements to be accepted 
as good photon candidates. Shower energies are required to be larger 
than 25~MeV in the EMC barrel region ($|\cos\theta|<0.8$) and larger than 50~MeV in 
the endcap ($0.86<|\cos\theta|<0.92$). The showers close to the boundary 
are poorly reconstructed and excluded from the analysis. To eliminate 
showers from charged particles, a photon must be separated from any 
charged tracks by more than $20^\circ$. Furthermore, in order to suppress 
electronic noise and energy deposits unrelated to the event, the EMC 
timing of the photon candidate is required to be in coincidence with 
the collision event, i.e., within 700 ns. 

The $K^0_S$ candidates are identified via the decay $K^0_S\to\pip\pim$. 
Secondary vertex fits are performed to all pairs of oppositely charged 
tracks in each event (assuming the tracks to be pions). The combination 
with the best fit quality is kept for further analysis if the invariant 
mass is within 10~\MeV~of the nominal $K^0_S$ mass~\cite{pdg}, and the 
decay length is more than twice the vertex resolution. The fitted $K^0_S$ 
information is used as an input for the subsequent kinematic fit.

In the $\psipp\to\gamma K^0_SK^\pm\pi^\mp$ channel selection, candidate events 
must contain at least four charged tracks and at least one good photon. 
After finding a $K^0_S$, the event should have exactly two additional 
charged tracks with zero net charge. A four-constraint (4C) kinematic fit 
is then applied to the selected final state with respect to the $\psipp$ 
four-momentum to reduce background and improve the mass resolution. 
The identification of the species of final state particles and the 
selection of the best photon when additional photons are found in an 
event are achieved by minimizing $\chi^2_{\rm total}=\chi^2_{\rm 4C}+\chi^2_{\rm PID}(K)+\chi^2_{\rm PID}(\pi)$ 
over all possible combinations, where $\chi^2_{\rm 4C}$ is the chi square 
of the 4C kinematic fit and $\chi^2_{\rm PID}(K)$ ($\chi^2_{\rm PID}(\pi)$)  
is the chi square of the PID for the kaon (pion). Events with $\chi^2_{\rm 4C}<20$ 
are accepted as $\gamma K^0_SK^\pm\pi^\mp$ candidates.

Background from  
$\psipp\to D^0\bar{D^0},~\bar{D}^0\to\pi^0K^0_S,~D^0\to\pi^+K^-$ or the charged conjugate process 
is removed by requiring $|M_{K^\pm\pi^\mp}-M_{D^0}|>3\sigma$, where $\sigma$ 
is the resolution of $M_{K^\pm\pi^\mp}$. To suppress background events 
with one additional photon, for instance $\pi^0 K^0_SK^\pm\pi^\mp$ events, the candidate events 
are also subjected to a 4C kinematic fit to the hypothesis $\gamma\gamma K^0_SK^\pm\pi^\mp$. 
We require the $\chi^2_{4C}$ of the $\gamma K^0_SK^\pm\pi^\mp$ hypothesis be less than 
that of the $\gamma\gamma K^0_SK^\pm\pi^\mp$ hypothesis.

\section{Data Analysis}
\label{sec:analysis}

By using large statistics MC samples we find the remaining dominant background can be classified into two categories: background from the continuum process $e^+e^-\to q\bar{q}$, which has smooth distributions around the $\etac$, $\etacp$ and $\chi_{c1}$ resonance; and background from the radiative tail of the $\psi(3686)$, which produces peaks within the signal regions ((2.90-3.05~\GeV) for $\etac$, (3.6-3.66~\GeV) for $\etacp$, and (3.49-3.54~\GeV) for $\chi_{c1}$). MC studies show that contributions from other known processes are negligible.

The background from the continuum process $e^+e^-\to q\bar{q}$ can be separated into three subcategories: events with an extra photon in the final state, $e^+e^-\to \pi^0 K^0_SK^\pm\pi^\mp$; events with the same final state as the signal, $e^+e^-\to \gamma_{\rm ISR}/\gamma_{\rm FSR} K^0_SK^\pm\pi^\mp$, where the photon comes from initial state radiation~(ISR) or final-state radiation~(FSR); and events with a fake photon candidate, $e^+e^-\to K^0_SK^\pm\pi^\mp$. 

Background from $e^+e^-\to \pi^0 K^0_SK^\pm\pi^\mp$, where a soft photon from $\pi^0\to\gamma\gamma$ is missing can be measured by reconstructing $e^+e^-\to \pi^0 K^0_SK^\pm\pi^\mp$ events from  data. The selection criteria are similar to those applied in the $\gamma K^0_SK^\pm\pi^\mp$ candidate selection but with an additional photon and a $\pi^0$ reconstructed from the selected photons.  A MC sample of $e^+e^-\to \pi^0 K^0_SK^\pm\pi^\mp$ is generated according to phase space to determine the relative efficiency of $\gamma K^0_SK^\pm\pi^\mp$ and $\pi^0 K^0_SK^\pm\pi^\mp$ selection criteria in each $M_{K^0_SK^\pm\pi^\mp}$ mass bin. By scaling the selected $\pi^0 K^0_SK^\pm\pi^\mp$ data sample with the efficiencies in each $M_{K^0_SK^\pm\pi^\mp}$ mass bin, we obtain the background contamination.

Background contributions from $e^+e^-\to (\gamma_{\rm ISR}/\gamma_{\rm FSR}) K^0_SK^\pm\pi^\mp$ are estimated with MC distributions for these processes normalized by the luminosity.  The generation of this sample includes the processes $e^+e^-\to {\rm hadrons}$  and $e^+e^-\to\gamma+{\rm hadrons}$, where the photon comes from ISR  or FSR (generated by {\sc photos}~\cite{Photos}) effects. The experimental Born cross section $\sigma(s)$ of $e^+e^-\to K^0_SK^\pm\pi^\mp$ obtained by the $BABAR$ Collaboration~\cite{babar} is used as input in the generator.

Background from the tail of the $\psi(3686)$ resonance production at $\sqrt{s} = 3.773$~GeV, including radiatively produced $\psi(3686)$ with soft ISR photon (i.e., $e^+e^-\to\gamma\psi(3686),~\psi(3686)\to \gamma X$ ($X$ stands for $\etac$, $\etacp$ or $\chi_{c1}$)), indistinguishable from the $\psipp$ decays, will produce peaks in the signal regions. Its contribution can be estimated by 
\begin{equation}\label{eqpsip}
N^b_{\psi(3686)}=\sigma(s)\times
\mathcal{L}\times\epsilon\times\Pi\mathcal{B}_i,
\end{equation}
where $\mathcal{L}$ is the integrated luminosity, $\epsilon$ is the detection efficiency for the final state in question, and $\mathcal{B}_i$ denotes the branching fraction for the intermediate resonance decays (i.e., $\mathcal{B}(\psi(3686)\to\gamma X)$, $\mathcal{B}(X\to K^0_SK^\pm\pi^\mp)$ and $\mathcal{B}(K^0_S\to\pi^+\pi^-)$). The cross section of $\psi(3686)$ production at $\sqrt{s} = 3.773$ GeV, $\sigma(s)$, can be expressed as 
\begin{eqnarray}\label{xsec_psip}
\sigma(s) & = & \int_{0}^{x_{\rm cut}}W(s,x)\cdot{\rm BW}(s^{\prime}(x))\cdot F_{X}(s^{\prime}(x))dx,
\end{eqnarray}
where $x$ is the scaled radiated energy in $e^+e^-\to\gamma_{\rm ISR}\psi(3686)$ ($x=2E_{\gamma_{\rm ISR}}/\sqrt{s}$); $s^{\prime}$ is the mass-squared with which the $\psi(3686)$ is produced ($s^{\prime}(x)=s(1-x)$); $W(s,x)$ is the ISR $\gamma$-emission probability~\cite{xsec-isrpsip}; ${\rm BW}(s^{\prime}(x))=12\pi\Gamma_R\Gamma_{ee}/[(s^{\prime}-M_R^2)^2+M_R^2\Gamma_R^2]$ is the relativistic Breit-Wigner formula describing the $\psi(3686)$ resonance; and $F_{X}(s^{\prime}(x))=(E_{\gamma}(s^{\prime})/E_{\gamma}(M_R^2))^3$ is the phase space factor between the $\psi(3686)$ produced with $\sqrt{s^{\prime}}$ mass and with its nominal mass, $M_R^2$, in which $E_{\gamma}$ is the energy of the transition photon in $\psi(3686)\to\gamma X$ decay. The $\psi(3686)$ nominal mass ($M_R$), total width ($\Gamma_R$) and $e^+e^-$ width ($\Gamma_{ee}$) are taken from the PDG. The threshold cutoff $x_{\rm cut}=1-m^2_X/s$ is chosen as the upper limit of integration in the definition of $\sigma(s)$, where $m_X$ is the nominal mass of $X$. The estimated numbers of background events are listed in Table~\ref{tab:psip_bkg}, where the errors arise dominantly from the uncertainties of the integrated luminosity, the cross section for $\psi(3686)$, the detection efficiencies, and the branching fractions.

\begin{table}[hbtp]
\begin{center}\small
\caption{The number of background events from the radiative tail of the $\psi(3686)$ resonance produced at $\sqrt{s} = 3.773$ GeV. The product branching fraction $\mathcal{B}(\psi(3686)\to\gamma\etacp,\etacp\to K^0_SK^\pm\pi^\mp)$ is taken from a previous \allowbreak{BESIII} measurement~\cite{prl-109-042003}, where the error is statistical only; others are taken from the PDG.}
\label{tab:psip_bkg}
\begin{tabular}{lcc}
\hline
\hline
X & $\mathcal{B}(\psi(3686)\to\gamma X\to\gamma K^0_SK^\pm\pi^\mp)$ & $N^b_{\psi(3686)}$ \\ \hline
    $\etac$ &  $(8.16\pm1.38)\times10^{-5}$ & $2.7\pm0.6$ \\ 
    $\etacp$ &  $(4.31\pm0.75)\times10^{-6}$ & $1.3\pm0.3$ \\ 
    $\chi_{c1}$ & $(3.36\pm0.31)\times10^{-4}$ & $19.8\pm3.1$ \\ 
\hline\hline
\end{tabular}
\end{center}
\end{table}

\begin{figure*}[hbtp]
\begin{center}\small
  {\includegraphics[width=0.47\linewidth]{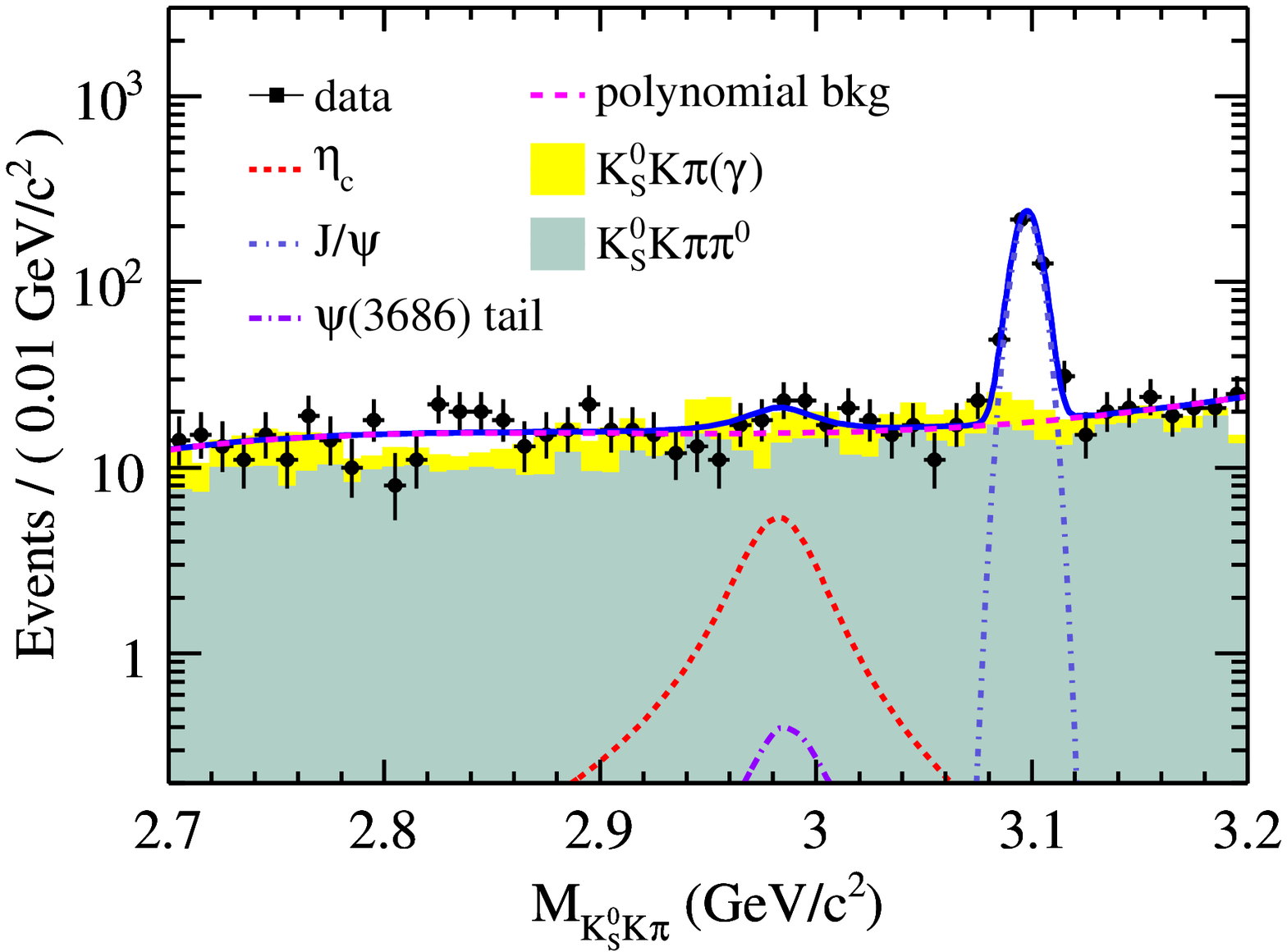}
  \put(-35,148) {\bf \footnotesize (a)}}
  {\includegraphics[width=0.47\linewidth]{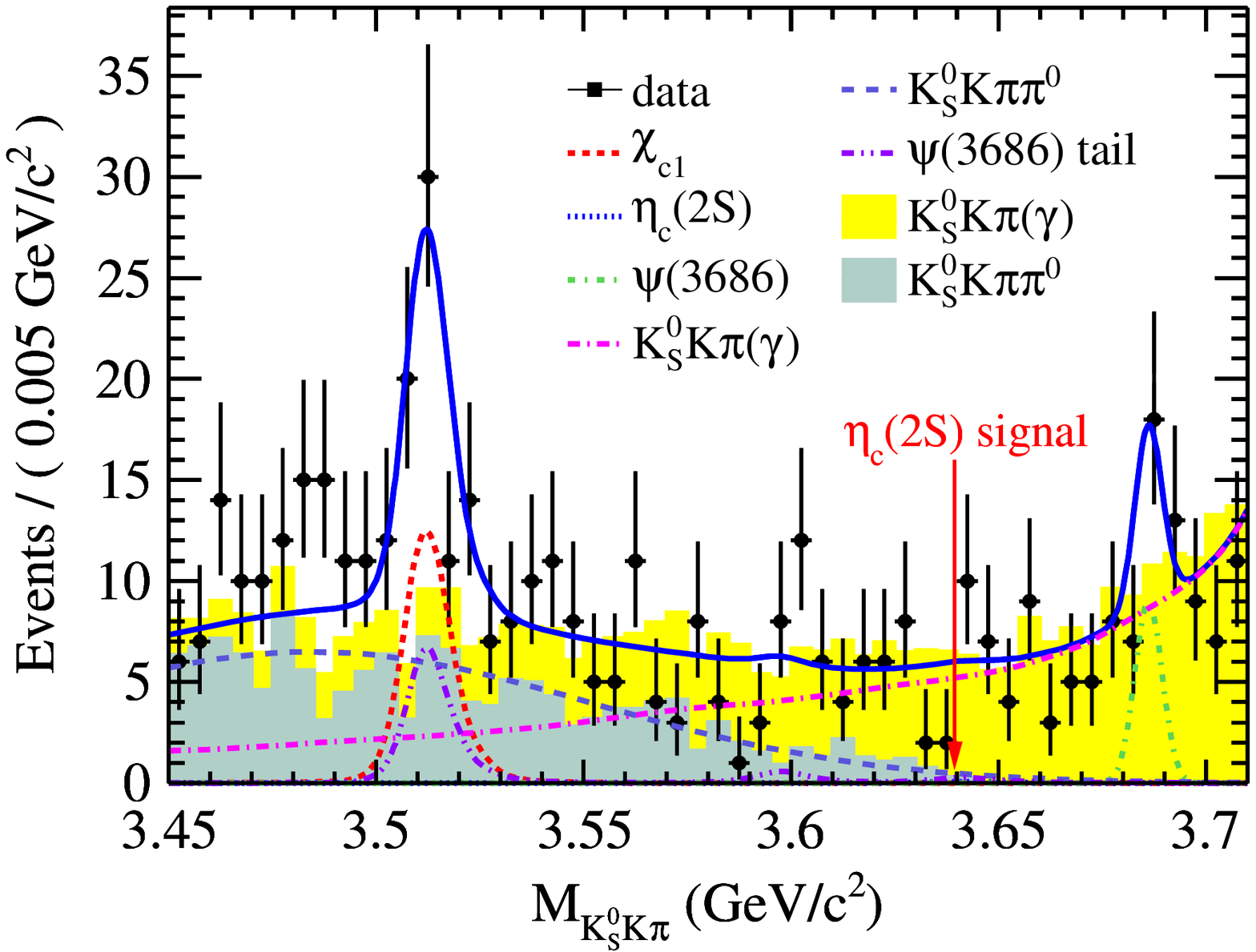}
  \put(-35,148) {\bf \footnotesize (b)}}
\caption{Invariant-mass spectrum for $K^0_SK^\pm\pi^\mp$ from data with  the estimated backgrounds and best-fit results superimposed in the (a)~$\etac$ and (b)~$\chi_{c1}$-$\etacp$ mass regions.  Dots with error bars are data. The shaded histograms represent the background contributions from $e^+e^-\to\pi^0K^0_SK^\pm\pi^\mp$ and  $e^+e^-\to(\gamma)K^0_SK^\pm\pi^\mp$, which are shown for comparison only. For the fitted curves, the solid lines show the total fit results. In (a), the $\etac$ and $\jpsi$ signals are shown as a short dashed line and a short dash-dotted line, respectively; the peaking background from the radiative tail of the $\psi(3686)$ is a long dash-dotted line; and the polynomial background is a long dashed line. In (b), the $\etacp$, $\chi_{c1}$ and  $\psi(3686)$ signals are shown as a dotted line (with too small amplitude but indicated by the arrow), a short dashed line, and a short dash-dotted line, respectively; the background from $e^+e^-\to(\gamma)K^0_SK^\pm\pi^\mp$ is a long dash-dotted line; the background from $e^+e^-\to\pi^0K^0_SK^\pm\pi^\mp$ is a long dashed line; and the peaking background from the radiative tail of the $\psi(3686)$ is a dash-dot-dotted line.}
\label{fit_kskpi}
\end{center}
\end{figure*}

Figure~\ref{fit_kskpi} shows the invariant-mass spectrum of $K^0_SK^\pm\pi^\mp$ for selected candidates, together with the estimated $e^+e^-\to\pi^0K^0_SK^\pm\pi^\mp$ and $e^+e^-\to(\gamma)K^0_SK^\pm\pi^\mp$ backgrounds. The estimated backgrounds can describe data well.  The summed background shapes from the continuum process $e^+e^-\to q\bar{q}$ are found to be flat in the $\etac$ mass region (2.7-3.2~\GeV) (Fig.~\ref{fit_kskpi}(a)) and smooth in the $\chi_{c1}$-$\etacp$ mass region (3.45-3.71~\GeV) (Fig.~\ref{fit_kskpi}(b)) without any enhancement in mass region of interest. 

The signal yields are extracted from unbinned maximum likelihood fits to the  distributions of $M_{K^0_SK^\pm\pi^\mp}$ in the $\etac$ and $\chi_{c1}$-$\etacp$ mass regions, separately, as shown in Figs.~\ref{fit_kskpi}(a) and \ref{fit_kskpi}(b), respectively. 

In the $\etac$ mass region, the fitting function consists of four components: the $\etac$ signal, ISR $\jpsi$, the peaking background from the radiative tail of the $\psi(3686)$, and the summed non-peaking background. The fitting probability density function~(PDF) as a function of mass~($m$) for the $\etac$  signal reads:
\begin{equation}\label{etac_pdf}
F(m)=G_{\rm res}\otimes(\epsilon(m)\times E^3_\gamma\times f_{\rm damp}(E_{\gamma})\times {\rm BW}(m)),
\end{equation}
where $G_{\rm res}$ is the experimental resolution function, $\epsilon(m)$ is the mass-dependent efficiency, $E_\gamma = \frac{m^2_{\psipp}-m^2}{2m_{\psipp}}$ is the energy of the transition photon in the rest frame of $\psipp$, $f_{\rm damp}(E_{\gamma})$ describes a factor to damp the diverging tail raised by $E^3_\gamma$ with  the functional form introduced by KEDR~\cite{KEDR}: 
\begin{equation}\label{damp_KEDR}
f_{\rm damp}^{\rm KEDR}=\frac{E^2_0}{E_\gamma E_0+(E_\gamma-E_0)^2},
\end{equation}
where $E_0 = \frac{m^2_{\psipp}-m^2_{\etac}}{2m_{\psipp}}$ is the peaking energy of the transition photon, and ${\rm BW}(m)$ is the  Breit-Wigner function with the resonance parameters of the $\etac$ fixed to the PDG. The mass-dependent efficiency is determined from MC simulation of the resonance decay according to the Dalitz plot distribution. The experimental resolution function, $G_{\rm res}$, is primarily determined from  a signal MC sample with the width of the resonance set to zero.  The consistency between data and MC simulation is checked by studying the process  $e^+e^-\to\gamma_{\rm ISR}\jpsi,~\jpsi\to K^0_SK^\pm\pi^\mp$. We use a smearing Gaussian function to describe the possible discrepancy between data and MC, whose parameters are determined by fitting the MC-determined $\jpsi$ shape convolved by this Gaussian function to the data. We assume that the discrepancy is  mass-independent. The line shape for the $\jpsi$ resonance is described by a Gaussian function with floating parameters. The shape of the peaking background from the radiative tail of the $\psi(3686)$ is obtained from the MC simulation with the amplitude fixed to the estimated number. We use a third-order Chebychev polynomial to represent the remaining flat background.

\begin{figure}[hbtp]
\begin{center}
  {\includegraphics[width=0.85\linewidth]{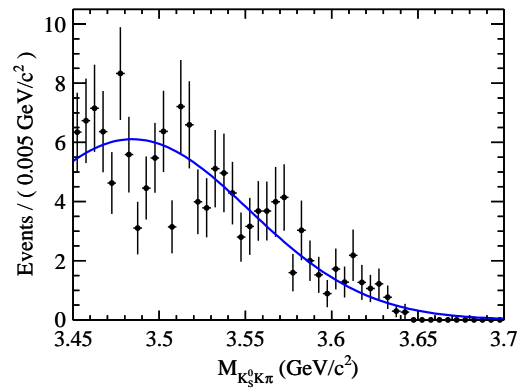}}
\caption{The measured background from $e^+e^-\to \pi^0 K^0_SK^\pm\pi^\mp$ (dots with error bars) with the expected size in the $\chi_{c1}$-$\etacp$ mass region.  The curve shows the fit with a Novosibirsk function.} \label{fit_bg_pi0_etacp}
\end{center}
\end{figure}

In the $\chi_{c1}$-$\etacp$ mass regions, the fitting function includes six  components: $\chi_{c1}$ and $\etacp$ signals; the $\psi(3686)$ peak; and backgrounds from the radiative tail of the $\psi(3686)$, $e^+e^-\to \pi^0 K^0_SK^\pm\pi^\mp$ and $e^+e^-\to (\gamma_{\rm ISR}/\gamma_{\rm FSR}) K^0_SK^\pm\pi^\mp$. The contribution from $\psipp\to\gamma\chi_{c2}\to\gamma K^0_SK^\pm\pi^\mp$, whose expected number of events is estimated to be less than 2.4 using a MC-determined detection efficiency and measured branching fractions~\cite{pdg}, is ignored in the fit. The line shapes for both the $\etacp$ and $\chi_{c1}$ resonances are also given by Eq.~\ref{etac_pdf}. The resonance parameters of the $\chi_{c1}$ and $\etacp$ are fixed to the PDG values. The line shape for  the $\psi(3686)$ resonance is described by a Gaussian function with its mean value fixed to that of the PDG. The background from the lower mass region is dominated by the  $e^+e^-\to\pi^0K^0_SK^\pm\pi^\mp$ process, which is studied in data as mentioned earlier. It is described by a Novosibirsk function~\cite{Novo} as shown in Fig.~\ref{fit_bg_pi0_etacp}. The determined  shape and magnitude of this background is fixed in the fit. The background on the higher mass region is $e^+e^-\to K^0_SK^\pm\pi^\mp(\gamma_{\rm ISR}/\gamma_{\rm FSR})$. We use the shape of the extracted $e^+e^-\to K^0_SK^\pm\pi^\mp(\gamma_{\rm ISR}/\gamma_{\rm FSR})$ MC sample to represent it, where the size is allowed to float. The shape of the peaking background from the radiative tail of the $\psi(3686)$ also comes from the MC simulation, and its magnitude is fixed to the expected number determined from the background study.

The results of the observed numbers of events  for the $\etac$, $\etacp$ and $\chi_{c1}$ are $29.3\pm18.2$, $0.4\pm8.5$ and $34.9\pm9.8$, respectively. The fits shown in Figs.~\ref{fit_kskpi}(a) and \ref{fit_kskpi}(b) have goodnesses of fit $\chi^2/ndf=27.1/42$ and $48.8/47$, which indicate reasonable fits. Since neither the $\etac$ nor the $\etacp$ signal is significant, we determine the upper limits on the number of signal events using the probability density function (PDF) for the expected number of signal events. The PDF is regarded as the likelihood distribution in fitting the invariant-mass spectrum in Fig.~\ref{fit_kskpi}(a) (Fig.~\ref{fit_kskpi}(b)) by setting the number of $\etac(\etacp)$ signal events from zero up to a very large number. The upper limit on the number of events at a 90\% confidence level (C.L.), $N_{\rm up}$, corresponds to  $\int_{0}^{N_{\rm up}}{\rm PDF}(x)dx/\int_{0}^{\infty}{\rm PDF}(x)dx=0.90$.

\section{Systematic uncertainties }
\label{sec:sys}

The systematic uncertainties of the branching fraction measurements mainly originate from the MDC tracking efficiency, photon detection, $K^0_S$ reconstruction, kinematic fitting, the $D^0$ and $\pi^0$ veto, $K^0_SK^\pm\pi^\mp$  intermediate states, the integrated luminosity of data, the cross section for $\psipp$, the damping function, and the fit to the invariant-mass distributions. The contributions are summarized in Table~\ref{tab:total_syserr} and  discussed in detail in the following paragraphs.

\begin{table}[hbtp]
\begin{center}\small
\caption{Summary of systematic uncertainties (\%) in the product 
branching fraction measurements of 
$\mathcal{B}(\psipp\to\gamma X)\times\mathcal{B}(X\to K^0_SK^\pm\pi^\mp)$,
where X stands for $\etac,~\etacp$, or $\chi_{c1}$.}
\label{tab:total_syserr}
\begin{tabular}{lccc}
\hline
\hline
Sources &  $\gamma\etac$ & $\gamma\etacp$ & $\gamma\chi_{c1}$\\ \hline
Tracking                                  &     2.0    &     2.0      & 2.0  \\
Photon reconstruction                     &     1.0    &     1.0      & 1.0  \\
$K^0_S$ reconstruction                    &     4.0    &     4.0      & 4.0  \\
Kinematic fitting                         &     3.9    &     5.5      & 5.3  \\
$D^0\&\pi^0$ veto                         &     3.2    &     3.2      & 3.2  \\
$K^0_SK^\pm\pi^\mp$ intermediate states   &     1.9    &     3.3      & 2.0  \\
$\mathcal{L}_{\psi(3770)}$                &     1.0    &     1.0      & 1.0  \\
$\sigma^0_{\psi(3770)}$                   &     7.8    &     7.8      & 7.8  \\
Fitting range                             &    \dots   &     8.1      & 3.2  \\
Non-peaking background                    &    \dots   &    10.2      & 8.9  \\
Background from $\psi(3686)$ tail         &    \dots   &     1.2      & 8.0  \\
Damping function                          &    \dots   &     1.9      & 0.3  \\
Mass and width of $\eta_c(2S)$            &    \dots   &    12.0      & \dots    \\
Total                                     &     10.6   &    21.3      & 16.7  \\
\hline
\hline
\end{tabular}
\end{center}
\end{table}

The difference in efficiency between data and MC simulation is 1\% for each $\pi^\pm$ or $K^\pm$ track that comes from the IP~\cite{k-eff,pi-eff}. So the uncertainty of the tracking efficiency is 2\%. The uncertainty due to photon reconstruction is estimated to be 1\% per photon~\cite{pho-eff}.

Three parts contribute to the complete efficiency for the $K^0_S$ reconstruction:  the geometric acceptance, the tracking efficiency, and the efficiency of  $K^0_S$ selection. The first part can be estimated using MC studies.  The other two are studied by the doubly tagged hadronic decay modes of $D^0\to K^0_S\pi^+\pi^-$ versus $\bar{D}^0\to K^+\pi^-$, $D^0\to K^0_S\pi^+\pi^-$ versus $\bar{D}^0\to K^+\pi^-\pi^0$, and $D^0\to K^0_S\pi^+\pi^-$ versus $\bar{D}^0\to K^+\pi^-\pi^-\pi^+$ and $\jpsi\to K^{*}\bar{K^0}+c.c.$. With these samples, the efficiency to reconstruct the $K_S^0$ from a pair of pions can be determined. The difference between data and MC, 4.0\%, is included in the systematic error.

\begin{figure}[hbtp]
\begin{center}
  {\includegraphics[width=0.85\linewidth]{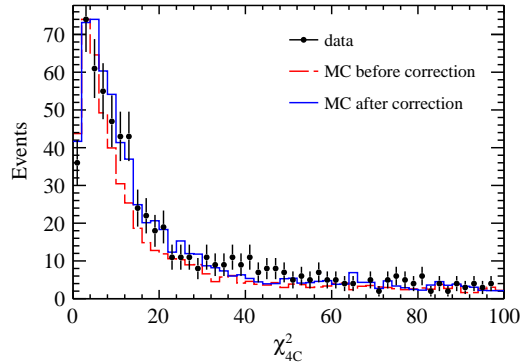}}
\caption{The comparison of $\chi_{4C}^2$ between data and MC for $e^+e^-\to\gamma_{\rm ISR}\jpsi,~\jpsi\to K^0_SK^\pm\pi^\mp$. The
dots with error bars are data, the dashed (solid) histogram represents MC simulation without (with) track parameter corrections.}
\label{chisq_4C_compare}
\end{center}
\end{figure}

There are differences between data and MC in the $\chi^2_{4C}$ distributions of the kinematic fit. These differences are dominantly due to the inconsistencies in the charged track parameters between data and MC.  We correct the track helix parameters ($\phi_0,\kappa,\tan\lambda$) to reduce the differences, where $\phi_0$ is the azimuthal angle specifying the pivot with respect to the helix center, $\kappa$ is  the reciprocal of the transverse momentum, and $\tan\lambda$ is the  slope of the track. The correction factors are extracted from pull distributions by using the control sample $\jpsi\to\phi f_0(980),~\phi\to K^+K^-,~f_0(980)\to\pi^+\pi^-$~\cite{cor-factor}.  The MC samples after correction are used to estimate the efficiency and to fit the invariant-mass spectrum. Figure~\ref{chisq_4C_compare} shows the $\chi_{4C}^2$ distributions before and after the corrections in MC and in data for the  control sample $e^+e^-\to\gamma_{\rm ISR}\jpsi,~\jpsi\to K^0_SK^\pm\pi^\mp$. The agreement between data and MC simulation does improve significantly after corrections, but differences still exist. The differences in the efficiencies, obtained using MC simulations with and without corrections, are taken as the systematic uncertainties as conservative estimations. These are 3.9\%, 5.5\% and 5.3\% for $\psipp\to\gamma\etac$, $\gamma\etacp$ and  $\gamma\chi_{c1}$, respectively.

The uncertainty due to the $D^0$ veto ($|m_{K^\pm\pi^\mp}-m_{D^0}|>3\sigma$) and the $\pi^0$ veto ($\chi_{4C}^2(\gamma K^0_SK^\pm\pi^\mp)<\chi_{4C}^2(\gamma\gamma K^0_SK^\pm\pi^\mp)$)  requirements is studied with a control sample of $e^+e^-\to\gamma_{\rm ISR}\jpsi,~\jpsi\to K^0_SK^\pm\pi^\mp$.  The total efficiency difference between data and MC is determined to  be 3.2\% for the $D^0$ and $\pi^0$ veto requirements together.

The reconstruction efficiencies are determined from MC simulations where $\etac,~\etacp,~\chi_{c1}\to K^0_SK^\pm\pi^\mp$ are generated according to the Dalitz plot distributions as described earlier. To estimate the uncertainties in the dynamics of the decays $\etac,\etacp,\chi_{c1}\to K^0_SK^\pm\pi^\mp$,  alternative MC samples treated as phase space distributions without any intermediate states are generated. The differences between efficiencies obtained with these two different generator models are taken as the systematic uncertainties due to possible intermediate states.

By analyzing Bhabha scattering events from the data taken at  $\sqrt{s} = 3.773$ GeV, the integrated luminosity of the data is measured to be 2.92~fb$^{-1}$, where the uncertainty is 1.0\%~\cite{lum}. To determine the total number of $\psipp$, we use the $\psipp$ Born-level cross section of at $\sqrt{s} = 3.773$~GeV, $\sigma^{0}_{\psipp} = (9.93\pm0.77)$~nb, which is calculated by a relativistic Breit-Wigner formula with the $\psipp$ resonance parameters~\cite{pdg}. The uncertainty  of $\sigma^0_{\psipp}$ is 7.8\%, arising dominantly  from the errors in the $\psipp$ resonance parameters.

The uncertainties from fitting the invariant-mass distributions of  $K^0_SK^\pm\pi^\mp$ are estimated by changing signal and background shapes and the corresponding fitting range. In the $\etac$ mass region, the fit-related uncertainties are obtained by varying the fitting range to [2.675, 3.225]~\GeV ~and [2.725, 3.175]~\GeV, changing the background function to a second-order polynomial, varying the parameters of  the $\etac$ by one standard deviation away from the PDG value, removing the damping factor, and changing the magnitude of the peaking background from the radiative $\psi(3686)$ tail by $\pm1\sigma$. The maximum $N_{\rm up}$ of the $\etac$  signal yield is used in the upper limit calculation.  In the $\chi_{c1}$-$\etacp$ mass region, the uncertainties due to the choice of fitting range are evaluated by varying the  range to [3.44, 3.72]~\GeV~and [3.46, 3.70]~\GeV. The largest differences in the results are assigned as errors. The uncertainties due to the choice of damping function is estimated from the  differences between results obtained with Eq.~\ref{damp_KEDR} and the form used by CLEO~\cite{damp-CLEO}: 
\begin{equation}\label{damp_CLEO}
f_{\rm damp}^{\rm CLEO}=\exp(-\frac{E_\gamma^2}{8\beta^2}),
\end{equation}
with $\beta = (65.0\pm2.5)$ MeV from CLEO's fit. The uncertainties caused by the parameters of the $\etacp$ are estimated by changing the mass and width values by $\pm1\sigma$. The background uncertainties dominantly come from the components $e^+e^-\to\pi^0K^0_SK^\pm\pi^\mp$ and the radiative tail of the $\psi(3686)$. We vary the shape parameters and magnitudes by $\pm1\sigma$, and take the differences on the results as systematic uncertainties.

The overall systematic uncertainties are obtained by combining all the sources of systematic uncertainties in quadrature, assuming they are independent.

\section{Results}

\begin{table*}[hbtp]
\begin{center}\small
\caption{The results for the branching fraction calculation. $\mathcal{B}_{\rm CLEO}(\psipp\to\gamma X)$ is the CLEO's measurement for the related branching fraction; $\Gamma(\psipp\to\gamma X)$ is the measured partial width for the related process calculated with $\Gamma(\psipp\to\gamma X)=\Gamma_{\psipp}\mathcal{B}(\psipp\to\gamma X)$;  $\Gamma_{\rm IML}$ and $\Gamma_{\rm LQCD}$ are the theoretical predictions of the partial width for $\psipp\to\gamma X$ based on IML and LQCD~\cite{LQCD} models, respectively. For the measured branching fractions, the first errors are statistical and the second ones are systematic.} 
\label{tab:results}
\begin{tabular}{lccc}
\hline
\hline
Quantity & $\etac$ & $\etacp$ & $\chi_{c1}$ \\ \hline
$N_{\rm obs}$ & $29.3\pm18.2$ & $0.4\pm8.5$ & $34.9\pm9.8$ \\ 
$N_{\rm up}$ & 56.8 & 16.1 & \dots \\
$\epsilon$ (\%) & 27.87 & 25.24 & 28.46 \\  
$\mathcal{B}(\psipp\to\gamma X\to\gamma K^0_SK^\pm\pi^\mp)$ ($\times10^{-6}$) 
& $<16$ & $<5.6$ & $8.51\pm2.39\pm1.42$ \\ 
$\mathcal{B}(\psipp\to\gamma X)$ ($\times10^{-3}$) 
& $<0.68$ & $<2.0$ & $2.33\pm0.65\pm0.43$ \\ 
$\mathcal{B}_{\rm CLEO}(\psipp\to\gamma X)$ ($\times10^{-3}$)
& \dots & \dots & $2.9\pm0.5\pm0.4$ \\ 
$\Gamma(\psipp\to\gamma X)$ (keV) & $<19$ & $<55$ & \dots \\
$\Gamma_{\rm IML}$ (keV) & $17.14^{+22.93}_{-12.03}$ & $1.82^{+1.95}_{-1.19}$ & \dots \\ 
$\Gamma_{\rm LQCD}$ (keV) & $10\pm11$ & \dots & \dots \\
\hline\hline
\end{tabular}
\end{center}
\end{table*}

We assume all the signal events from the fit come from resonances  ($\etac$, $\etacp$, $\chi_{c1}$), neglecting possible interference between the signals and non-resonant contributions. The upper limits on the product branching fractions $\mathcal{B}(\psipp\to\gamma\etac(\etacp) \to\gamma K^0_SK^\pm\pi^\mp)$ are calculated with
\begin{eqnarray}
\mathcal{B}(\psipp\to\gamma\etac(\etacp)\to\gamma K^0_SK^\pm\pi^\mp) 
\nonumber\\ 
<\frac{N_{\rm up}/(1-\sigma_{\rm syst.})}{\epsilon\cdot\mathcal{L}\cdot\sigma^0_{\psipp}\cdot(1+\delta)\cdot\mathcal{B}(K^0_S\to\pi^+\pi^-)},
\label{eq:upperlimit}
\end{eqnarray}
where $N_{\rm up}$ is the upper limit number on the signal size, $\sigma_{\rm syst.}$  is the total systematic error, $\epsilon$ is the efficiency of the event selection, $\mathcal{L}$ is the integrated  luminosity of the data, $\sigma^0_{\psipp}$ is the Born-level cross section for the $\psipp$  produced at 3.773~GeV, $(1+\delta)=0.718$ is the radiative correction factor, obtained from the {\sc kkmc} generator with the $\psipp$ resonance parameters~\cite{pdg} as input,  and $\mathcal{B}(K^0_S\to\pi^+\pi^-)$ is the branching  ratio for $K^0_S\to\pi^+\pi^-$.  The product branching fraction $\mathcal{B}(\psipp\to\gamma\chi_{c1} \to\gamma K^0_SK^\pm\pi^\mp)$ is derived from  
\begin{eqnarray}
\mathcal{B}(\psipp\to\gamma\chi_{c1}\to\gamma K^0_SK^\pm\pi^\mp) 
 =  \nonumber \\ 
\frac{N_{\rm obs}}{\epsilon\cdot\mathcal{L}\cdot\sigma^0_{\psipp}
\cdot(1+\delta)\cdot\mathcal{B}(K^0_S\to\pi^+\pi^-)}, 
\end{eqnarray}
where $N_{\rm obs}$ is the observed number of events from the fit and others are the same as described in Eq.~\ref{eq:upperlimit}.

Dividing these product branching fractions by  $\mathcal{B}(\etac\to K^0_SK^\pm\pi^\mp)=\frac{1}{3}\mathcal{B}(\etac\to K\bar{K}\pi)=\frac{1}{3}(7.2\pm0.6)\%$, $\mathcal{B}(\etacp\to K^0_SK^\pm\pi^\mp)=\frac{1}{3}\mathcal{B}(\etacp\to K\bar{K}\pi)=\frac{1}{3}(1.9\pm1.2)\%$ and $\mathcal{B}(\chi_{c1}\to K^0_SK^\pm\pi^\mp)=(3.65\pm0.30)\times10^{-3}$ from the PDG, we obtain $\mathcal{B}_{\rm up}(\psipp\to\gamma\etac)$ and  $\mathcal{B}_{\rm up}(\psipp\to\gamma\etacp)$ at a 90\% C.L. and  $\mathcal{B}(\psipp\to\gamma\chi_{c1})$. All the results are summarized in Table~\ref{tab:results}.

\section{Summary}

In summary, using the 2.92~fb$^{-1}$ data sample taken at $\sqrt{s} = 3.773$~GeV  with the BESIII detector at the BEPCII collider, searches for the radiative  transitions between the $\psipp$ and the $\etac$ and the $\etacp$ through the decay process  $\psipp\to\gamma K^0_SK^\pm\pi^\mp$ are presented. No significant $\etac$ and $\etacp$ signals are observed. We set upper limits on the branching fractions at a 90\% C.L. 
\begin{eqnarray}
&\mathcal{B}(\psipp\!\to\!\gamma\etac\!\to\!\gamma K^0_SK^\pm\pi^\mp)\!<\!1.6\!\times\!10^{-5},\\
&\mathcal{B}(\psipp\!\to\!\gamma\etacp\!\to\!\gamma K^0_SK^\pm\pi^\mp)\!<\!5.6\!\times\!10^{-6},\\
&\mathcal{B}(\psipp\!\to\!\gamma\etac)\!<\!6.8\!\times\!10^{-4},\\
&\mathcal{B}(\psipp\!\to\!\gamma\etacp)\!<\!2.0\!\times\!10^{-3}.
\end{eqnarray}
We also report
\begin{eqnarray}
&\mathcal{B}(\psipp\!\to\!\gamma\chi_{c1}\!\to\!\gamma K^0_SK^\pm\pi^\mp)
\nonumber\\
&\!=\!(8.51\pm2.39\pm1.42)\!\times\!10^{-6},\\
&\mathcal{B}(\psipp\!\to\!\gamma\chi_{c1})
\!=\!(2.33\pm0.65\pm0.43)\!\times\!10^{-3},
\end{eqnarray}
where the first errors are statistical and the second ones are systematic.

Table~\ref{tab:results} compares the results of our measurements with the theoretical predictions from IML~\cite{IML} and lattice QCD~\cite{LQCD}  calculations, as well as those of CLEO\cite{gchi2}, if any. The upper limit for $\Gamma(\psipp\to\gamma\etac)$ is just within the error range of the theoretical predictions. However, the upper limit for $\Gamma(\psipp\to\gamma\etacp)$ is much larger than the prediction and is limited by statistics and the dominant systematic error, which stems from the uncertainty in the branching fraction of $\etacp\to K^0_SK^\pm\pi^\mp$. The measured branching fraction for $\psipp\to\gamma\chi_{c1}$ is consistent with the CLEO result, but the small branching ratio for $\chi_{c1}\to K^0_SK^\pm\pi^\mp$ reduces our sensitivity so that the precision is inferior to that of CLEO, which used four high-branching-fraction decays to all-charged hadronic final states ($2K$, $4\pi$, $2K2\pi$, and $6\pi$).

\begin{acknowledgements}
The BESIII collaboration thanks the staff of BEPCII and the computing 
center for their strong support. This work is supported in part by the 
Ministry of Science and Technology of China under Contract No. 2009CB825200; 
Joint Funds of the National Natural Science Foundation of China under 
Contracts Nos. 11079008, 11179007, 11179014, 11179020, U1332201; National Natural Science 
Foundation of China (NSFC) under Contracts Nos. 10625524, 10821063, 10825524, 
10835001, 10935007, 11125525, 11235011; the Chinese Academy of Sciences (CAS) 
Large-Scale Scientific Facility Program; CAS under Contracts Nos. KJCX2-YW-N29, 
KJCX2-YW-N45; 100 Talents Program of CAS; German Research Foundation DFG 
under Contract No. Collaborative Research Center CRC-1044; Istituto Nazionale 
di Fisica Nucleare, Italy; Ministry of Development of Turkey under Contract 
No. DPT2006K-120470; U. S. Department of Energy under Contracts Nos. 
DE-FG02-04ER41291, DE-FG02-05ER41374, DE-FG02-94ER40823, DESC0010118; 
U.S. National Science Foundation; University of Groningen (RuG) and the 
Helmholtzzentrum fuer Schwerionenforschung GmbH (GSI), Darmstadt; WCU Program 
of National Research Foundation of Korea under Contract No. R32-2008-000-10155-0.
\end{acknowledgements}


\begin{thebibliography}{99}
\bibitem{Rosner} J.~L.~Rosner, Phys. Rev. D {\bf 64}, 094002 (2001).
\bibitem{prl39} P.~A.~Rapidis {\it et al.}, Phys. Rev. Lett. {\bf 39}, 526 (1977).
\bibitem{prl40} W.~Bacino {\it et al.}, Phys. Rev. Lett. {\bf 40}, 671 (1978).
\bibitem{Abli1} M.~Ablikim {\it et al.} (BES Collaboration), Phys. Rev. D {\bf 76}, 122002 (2007).
\bibitem{Abli2} M.~Ablikim {\it et al.} (BES Collaboration), Phys. Lett. B {\bf 659}, 74 (2008).
\bibitem{Abli3} M.~Ablikim {\it et al.} (BES Collaboration), Phys. Rev. Lett. {\bf 97}, 121801 (2006).
\bibitem{Abli4} M.~Ablikim {\it et al.} (BES Collaboration), Phys. Lett. B {\bf 641}, 145 (2006). 
\bibitem{CLEO-nonDD} D.~Besson {\it et al.} (CLEO Collaboration), Phys. Rev. Lett. {\bf 104}, 159901(E) (2010).
\bibitem{pipijpsi} J.~Z.~Bai {\it et al.} (BES Collaboration), Phys. Lett. B {\bf 605}, 63 (2005).
\bibitem{nondd-cleo} N.~E.~Adam {\it et al.} (CLEO Collaboration), Phys. Rev. Lett. {\bf 96}, 082004 (2006).
\bibitem{gchi1} T.~E.~Coan {\it et al.} (CLEO Collaboration), Phys. Rev. Lett. {\bf 96}, 182002 (2006). 
\bibitem{gchi2} R.~A.~Briere {\it et al.} (CLEO Collaboration), Phys. Rev. D {\bf 74}, 031106 (2006). 
\bibitem{phieta} G.~S.~Adams {\it et al.} (CLEO Collaboration), Phys. Rev. D {\bf 73}, 012002 (2006).
\bibitem{pdg} J.~Beringer {\it et al.} (Particle Data Group), Phys. Rev. D {\bf 86}, 010001 (2012).
\bibitem{IML} G.~Li and Q.~Zhao, Phys. Rev. D {\bf 84}, 074005 (2011).
\bibitem{besnim} M.~Ablikim {\it et al.} (BESIII Collaboration), Nucl. Instrum. Methods Phys. Res., Sect. A {\bf 614}, 345 (2010).
\bibitem{taucharm} D.~M.~Asner {\it et al.}, Int. J. Mod. Phys. A {\bf 24}, 499 (2009).
\bibitem{GEANT4} S.~Agostinelli {\it et al.} (GEANT4 Collaboration), Nucl. Instrum. Methods Phys. Res.,
Sect. A {\bf 506}, 250 (2003).
\bibitem{BOOST} Z.~Y.~Deng {\it et al.}, Chinese Phys. C {\bf 30}, 371 (2006).
\bibitem{kkmc1} S.~Jadach, B.~F.~L. Ward, and Z.~Was, Comput. Phys. Commun. {\bf 130}, 260 (2000).
\bibitem{kkmc2} S.~Jadach, B.~F.~L. Ward, and Z.~Was, Phys. Rev. D {\bf 63}, 113009 (2001).
\bibitem{Belle-etacp} A.~Vinokurova {\it et al.} (Belle Collaboration), Phys. Lett. B {\bf 706}, 139 (2011).
\bibitem{evtgen} D.~J.~Lange, Nucl. Instrum. Methods Phys. Res., Sect. A {\bf 462}, 152 (2001).
\bibitem{lund} J.~C.~Chen, G.~S.~Huang, X.~R.~Qi, D.~H.~Zhang, and Y.~S.~Zhu, Phys.
Rev. D {\bf 62}, 034003 (2000).
\bibitem{Photos} E.~Barberio and Z.~Was, Comput. Phys. Commun. {\bf 79}, 291 (1994).
\bibitem{babar} B.~Aubert {\it et al.} ($BABAR$ Collaboration), Phys. Rev. D {\bf 77}, 092002 (2008).
\bibitem{xsec-isrpsip} M.~Benayoun {\it et al.} Mod. Phys. Lett. A {\bf 14}, 2605 (1999).
\bibitem{prl-109-042003} M.~Ablikim {\it et al.} (BESIII Collaboration), Phys. Rev. Lett. {\bf 109}, 042003 (2012).
\bibitem{KEDR} V.~V.~Anashin, Int. J. Mod. Phys. Conf. Ser. {\bf 02}, 188 (2011).
\bibitem{Novo}
The Novosibirsk function is defined as $f(m_{ES})=A_{S}{\rm exp}(-0.5\ln^{2}[1+\Lambda\tau\cdot(m_{ES}-m_0)]/\tau^2+\tau^2)$, 
where $\Lambda=\sinh(\tau\sqrt{\ln4})/(\sigma\tau\sqrt{\ln4})$, the peak position is $m_0$, the width is $\sigma$, 
and $\tau$ is the tail parameter.
\bibitem{k-eff} M.~Ablikim {\it et al.} (BESIII Collaboration), Phys. Rev. Lett. {\bf 107}, 092001 (2011).
\bibitem{pi-eff} M.~Ablikim {\it et al.} (BESIII Collaboration), Phys. Rev. D {\bf 83}, 112005 (2011).
\bibitem{pho-eff} M.~Ablikim {\it et al.} (BESIII Collaboration), Phys. Rev. D {\bf 81}, 052005 (2010).
\bibitem{cor-factor} M.~Ablikim {\it et al.} (BESIII Collaboration), Phys. Rev. D {\bf 87}, 052005 (2013).
\bibitem{lum} M.~Ablikim {\it et al.} (BESIII Collaboration), Chinese Phys. C {\bf 37}, 123001 (2013).
\bibitem{damp-CLEO} R.~E.~Mitchell {\it et al.} (CLEO Collaboration), Phys. Rev. Lett. {\bf 102}, 011801 (2009); {\bf 106}, 159903(E) (2001).
\bibitem{LQCD} J.~J.~Dudek, R.~Edwards, and C.~E.~Thomas, Phys. Rev. D {\bf 79}, 094504 (2009).
\end{thebibliography}
\end{document}